\documentclass[a4paper,aps,prb,twocolumn,floatfix,showpacs,amsfonts,amssymb,superscriptaddress,amsmath]{revtex4} 
\usepackage[T1]{fontenc}
\usepackage[latin1]{inputenc}
\usepackage{graphics}
\usepackage{amssymb}
\usepackage{amsmath}
\usepackage{overpic}

\newcommand{\be}{\begin{equation}}
\newcommand{\ee}{\end{equation}}
\newcommand{\bea}{\begin{eqnarray}}
\newcommand{\eea}{\end{eqnarray}}

\newcommand{\bkfa}{Ba$_{1-x}$K$_{x}$Fe$_{2}$As$_{2}$}

\def\a{\alpha}
\def\b{\beta}
\def\e{\varepsilon}

\def\g{\gamma}
\def\m{\mu}

\def\o{\omega}

\def\G{\Gamma}
\def\D{\Delta}

\def\ra{\rightarrow}

\def\Ra{\Rightarrow}

\def\bk{{\bf k}}

\def\OO{{\cal{O}}}

\def\nn{\nonumber}
\def\lb{\label}
\def\pref#1{(\ref{#1})}

\newcount\bozza \bozza=0
\ifnum\bozza=1
\newdimen\shift \shift=-2truecm
\def\lb#1{%
{\label{#1}\rlap{\kern\shift{$\scriptstyle#1$}}}}
\else\def\lb#1{\label{#1}} \fi

\begin{document} 
\title{Spectroscopic and thermodynamic properties in a four-band model 
for pnictides}

\author{L. Benfatto}
\affiliation{Centro Studi e Ricerche ``Enrico Fermi'', via Panisperna 89/A, 00184, Rome, Italy} 

\affiliation{SMC Research Center, CNR-INFM, and ISC-CNR,
via dei Taurini 19, 00185 Rome, Italy}

\affiliation{Dipartimento di Fisica, Universit\`a ``La Sapienza'',
Piazzale A. Moro 2, 00185 Rome, Italy}

\author{E. Cappelluti}
\affiliation{SMC Research Center, CNR-INFM, and ISC-CNR,
via dei Taurini 19, 00185 Rome, Italy}
\affiliation{Dipartimento di Fisica, Universit\`a ``La Sapienza'',
Piazzale A. Moro 2, 00185 Rome, Italy}

\author{C. Castellani}
\affiliation{Dipartimento di Fisica, Universit\`a ``La Sapienza'',
Piazzale A. Moro 2, 00185 Rome, Italy}
\affiliation{SMC Research Center, CNR-INFM, and ISC-CNR,
via dei Taurini 19, 00185 Rome, Italy}

\begin{abstract}
In this paper we provide a comprehesive analysis of different properties of
pnictides both in the normal and superconducting state, with a particular
focus on the optimally-doped Ba$_{1-x}$K$_{x}$Fe$_2$As$_2$ system.  We show
that, by using the band dispersions experimentally measured by ARPES, a
four-band Eliashberg model in the intermediate-coupling regime can account
for both the measured hierarchy of the gaps and for several spectroscopic
and thermodynamic signatures of low-energy renormalization. These include
the kinks in the band dispersion and the effective masses determined via
specific-heat and superfluid-density measurements. We also show that,
although an intermediate-coupling Eliashberg approach is needed to account
for the magnitude of the gaps, the temperature behavior of the
thermodynamic quantities does not show in this regime a significant
deviation with respect to weak-coupling BCS calculations. This can explain
the apparent success of two-band BCS fits of experimental data reported
often in the literature.
\end{abstract}
\pacs{74.20.-z, 74.25.Jb, 74.25.Bt} 
 
\date{\today} 
\maketitle 

\section{Introduction}

The recent discovery of superconductivity in pnictides\cite{kamihara}
prompted an intense experimental and theoretical research about the
properties of these materials. At the very beginning the analogies between
pnictides and cuprate superconductors (e.g. the layered structure and the
phase diagram) suggested that a similar route to high-temperature pairing
could be at play in these two classes of materials.\cite{norman,sadovsky}
However, a large experimental evidence has been accumulated so far that
significant differences between pnictides and cuprates are also important,
starting from the very basic fact that pnictides have a multiband
structure. According to LDA calculations, indeed, the band structure of
pnictides near the Fermi level is characterized by two hole-like bands
around the $\Gamma$ point, and two electron-like bands around the M points
of the reduced Brillouin zone.\cite{mazin,singh_bkfa,ma_lda_bkfa} A third
hole-like band at the $\G$ point could be expected to cross the Fermi level
in some materials, but eventually it moves below the Fermi level when the
experimental value of the apical As position is used in LDA
calculations.\cite{boeri_apical}

Despite the large theoretical work devoted to address the outcomes of
multiband superconductivity, many open issues still remain about a direct
comparison between the theoretical predictions and the experiments, or
between the outcomes of different experimental probes. A first issue
concerns the experimental observation of only two gap values
in hole-doped 122 compounds,\cite{ding1,ding2} whereas
in a  multiband BCS approach one would generically expect
a different gap in each band,
depending on the coupling and on the density of states (DOS) of the several
pockets involved in the pairing.\cite{barzykin,benfatto_4bands,dolgov}
This is true in particular for
hole-doped {\bkfa}(BKFA), where many detailed experimental findings have
been accumulating due to the existence of large crystals. Here
angle-resolved photo-emission spectroscopy (ARPES) has reported quite
different DOS in the hole and electron pockets involved in the
pairing,\cite{ding2} so that at the BCS level one could expect to observe three
gaps, one for each hole band and one for the (almost degenerate) electron
bands. On the contrary,
ARPES experiments have reported only two different gaps: a
large one on the inner hole pocket and on the electron ones
(with $\D/T_c\sim 3.5$), 
and a small one (with $\D/T_c\sim 1.8$) on the outer hole
pocket.\cite{ding1,ding2}

Besides the non-BCS hierarchy of the gaps, further difficulties arise in
the attempt to reconcile ARPES data with several thermodynamic
measurements. For instance, photoemission experiments performed by several
groups in different pnictide materials have shown that there is a
substantial renormalization of the whole band structure with respect to LDA
predictions, with a reduction at least of a factor
two.\cite{ding2,lu,yang,yi} At the same time, the estimates of the specific
heat coefficient $C_V/T$ obtained by using the ARPES bandwidth, despite
being substantially larger than LDA, are still about a factor two smaller
than the values measured in the normal state for 122
compounds.\cite{canfield_prb08,mu,loid_cm09} This comparison calls for a
dichotomy between high-energy and low-energy mass renormalization, that
must be accounted for by different mechanisms.  Recently, a similar
distinction between renormalization effects operative at different energy
scales has been pointed out also in optical-conductivity measurements in
1111 compounds.\cite{basov_sumrule_09}

As far as the temperature dependence of the specific heat in the
superconducting state is concerned, the comparison with ARPES is again
compelling: indeed, despite the large $\Delta/T_c$ values reported by ARPES
that call for an intermediate/strong coupling pairing mechanism, the
temperature profile of the specific heat can be well reproduced by a simple
BCS fit.\cite{mu,loid_cm09} A similar result arises from the analysis of
superfluid-density
measurements,\cite{matsuda,hiraishi,uemura_low,uemura_high,li,
khasanov,prozorov} where two-gaps BCS fits seem to work quite well once
that the experimental $\D/T_c$ ratios are implemented.

In this paper we provide a systematic analysis of spectral and
thermodynamic properties in pnictides with the goal of reconciling the
results obtained with the different probes. Our analysis is based on a
four-band model where carriers interact with bosonic excitations treated
within the Eliashberg approach. As we shall see, the observed hierarchy of
the gaps calls for a predominant interband nature of the interactions,
making spin fluctuations the most natural candidates for the pairing
glue.\cite{arita,chubukov} We shall focus in particular on the effects of
the exchange of spin fluctuations on several spectroscopic and
thermodynamical properties.  Since the typical energy scale for spin
fluctuations is of the order of 20
meV,\cite{matan,osborn,christianson,bourges_spin} it cannot be responsible
for the overall band narrowing observed by ARPES, that is operative up to a
rather high energy. In this context we shall extract input band parameters
directly from the high-energy ARPES measurements, by assuming that their
renormalization with respect to LDA calculations originates from electronic
correlations.\cite{georges,vollhardt,yang2} This approach is thus different
from what discussed in Ref.\ [\onlinecite{benfatto_4bands}], when the
experimental ARPES determination of the bands for doped BKFA was not yet
available. In that case it was shown that, within a BCS approach, the
experimental observation of only two gap values could be accounted for by a
suitable (moderate) renormalization of the LDA band parameters. Here
instead we show that the experimental measurement of the band structure
together with the gap values on each band provide in this material a
compelling constraint for the microscopic theory.  Within this framework we
estimate the magnitude of the different interband couplings from a
comparison with the measured gaps. We find that the dimensionless couplings
vary from $\lambda \simeq 0.2$ to $\lambda \simeq 1.6$, depending on the
band.  We also calculate the additional mass renormalization due to the
exchange of spin fluctuations.  These low-energy features are hardly
visible in ARPES but they are responsible for the large effective mass of
the charge carriers probed by specific-heat measurements, that are sensible
to excitations near the Fermi level, solving then the apparent
contradiction between the different set of measurements.  Finally, we
analyze the temperature dependence of the specific heat and superfluid
density, and we show that at the coupling values relevant for pnictides we
do not observe significant deviations from a conventional BCS profile,
explaining the apparent success of the BCS fits proposed in the literature.

It is worth pointing out the differences between our approach and previous
works on multiband Eliashberg calculations proposed in the
literature.\cite{dolgov,umma} The tendency of the gaps to assume the same
value in strongly nested bands within the Eliashberg theory was already
noted in Ref.\ [\onlinecite{dolgov}].  However, the authors considered
there a two-band model, so that it was impossible to reproduce the second
smaller gap value measured by ARPES, which is realized in a third, less
coupled band. Indeed, a correct approach to pnictides requires using at
least a four-band model with an anisotropic interband pairing, as it was
pointed out previously within a BCS scheme in Ref.\
[\onlinecite{benfatto_4bands}]. An Eliashberg approach to a four-band model
has been explored recently in Ref.\ [\onlinecite{umma}], where the authors
were aimed to reproduce exactly the experimental ratios $\D/T_c$ in the
various bands.  An extremely large coupling $\lambda > 4$ was there found
for BKFA.  Such analysis disregards however the fact that an accurate
estimate of $T_c$ within the mean-field-like Eliashberg theory is doubtful
in these almost two-dimensional materials, where superconducting
fluctuations are expected to be relevant due to the low
dimensionality,\cite{varlamov, benfatto_fluct} leading to a lowering of the
real $T_c$ in comparison with the mean-field estimate. In this situation we
prefer to concentrate our analysis on the consistency between the gap
values and the density of states. As mentioned above, assuming a typical
energy scale $\omega_0\approx 20$ meV for the characteristic
spin-fluctuations, we get $\lambda_i \simeq 0.2-1.6$, much lower than in
Ref.\ [\onlinecite{umma}] ($i$ being here the band index). On the other
hand, as we shall show below, these values appear to be perfectly
compatible with the thermodynamical properties, whereas stronger couplings
would be inappropriate, because the low-energy renormalization of the
charge carriers would be too large compared to the experimental outcomes
from specific-heat and superfluid-density measurements.  Our estimates of
the interband coupling $\lambda_i \simeq 0.2-1.6$ in BKFA locate this
material in the weak-intermediate coupling regime.  This observation could
suggest that analytical expressions {\em \`a la}
McMillan-Allen-Dynes\cite{carbotte} would be appropriate, as proposed in
Ref.\ [\onlinecite{dolgov}]. This is however not the case in multiband
systems where, as we show below, McMillan-Allen-Dynes-like expressions
can qualitative fail already above very weak coupling $\lambda \gtrsim 0.2$,
so that a numerical solution of the multiband Eliashberg equation is
required.

The structure of the paper is the following. In Section II, we briefly
review the results of a two-band model in order to elucidate the
differences between the BCS and Eliashberg approach and the need of a
numerical solution even in the weak/intermediate coupling regime. The
reader interested only in the comparison with the experiments can skip this
technical discussion, and refer directly to Section
III, where we introduce the full four-bands model, and we show that at
intermediate coupling the Eliashberg theory can reproduce the experimentally
measured gap values in pnictides. In Section IV we show the results for the
specific heat and the superfluid density. Finally, in Section V we draw
some conclusions and we discuss the perspectives of our work.

\section{Two-band model}

The BCS theory is characterized by a number of universal behaviors (as the
$T_c$ vs. $\lambda$ relation, the $\Delta/T_c=1.76$ ratio, etc.) which are
strictly valid only in the limit where the dimensionless coupling $\lambda
\rightarrow 0$.  Deviations from these universal results are related to
intermediate/strong coupling effects, so that the analysis of such
deviations could be employed in principle to estimate the strength of the
coupling $\lambda$.  McMillan-Allen-Dynes-like formulas,\cite{carbotte}
based on a controlled expansion in power of $\lambda$, could be quite
useful in this context because in single-band models they provide
analytical expressions to quantify these effects without resorting to a
numerical solution of the Eliashberg equations.

In this Section we show that in the multiband case with predominant
interband interaction the situation is quite different. Indeed, the
McMillan-Allen-Dynes-like expansion reproduces the Eliashberg behavior as
function of $\lambda$ only for very weak coupling $\lambda \lesssim 0.2$,
whereas for larger couplings a full numerical solution of the Eliashberg
equations is required.  We demonstrate this result for simplicity within a
two-band system, previously addressed in Ref.\ [\onlinecite{dolgov}].  We
thus write the general Eliashberg equations for a purely interband
interaction which is taken to be repulsive in the Cooper channel:
\bea
Z_1(n)\Delta_1(n)
&=&
-\lambda_{12}\pi T
\sum_m D(n-m)
%\nonumber\\
%&&\times
\frac{\Delta_2(m)}{\sqrt{\omega_m^2+\Delta_2^2(m)}},\nn\\
\lb{d1el}\\
Z_1(n)
&=&
1+\lambda_{12}\frac{\pi T}{\omega_n}
\sum_m D(n-m)
%\nonumber\\
%&&\times
\frac{\omega_m}{\sqrt{\omega_m^2+\Delta_2^2(m)}},\nn\\
\lb{z1el}
\eea
together with an  equivalent set of equations for $\D_2, Z_2$
related through a coupling constant $\lambda_{21}$.
Here for sake of shortness we denote
the dependence on the Matsubara frequency $\omega_n$ 
of the gap function $\D_i$ and the the renormalization
function $Z_i$ for the band $i$  as 
$\Delta_i(n)=\Delta_i(i\omega_n)$ and $Z_i(n)=Z_i(i\omega_n)$.
$D(n-m)=D(\o_n-\o_m)$ is the
boson propagator, which is related to the
Eliashberg spectral function $B(\Omega)$ by the relation
$D(n-m)=\int 2\Omega d\Omega B(\Omega)/[(\o_n-\o_m)^2+\Omega^2]$.
The dimensionless coupling constants $\lambda_{12}$, $\lambda_{21}$ can be
expressed in term of an unique energy coupling $G>0$ weighted
by the appropriate density of states $N_i$, namely
$\lambda_{12}=GN_2$, $\lambda_{21}=GN_1$.

Eqs. (\ref{d1el})-(\ref{z1el}) can be solved self-consistently to obtain a
numerical exact solution of the Eliashberg equations, assuming, for
simplicity, an Einstein boson spectrum
$B(\Omega)=(\omega_0/2)\delta(\Omega-\omega_0)$, where $\omega_0$
is the characteristic boson energy. For a repulsive
interaction, $G>0$, the gaps in the two bands have opposite signs, so that
the order parameter has a $s_\pm$ symmetry. In the rest of the Section we
will assume conventionally $\D_1>0$ and $\D_2<0$. Moreover, to make a
direct comparison with Ref.\ [\onlinecite{dolgov}], we consider the case
where the ratio $B=N_2/N_1$ of the DOS in the two bands is $B=2.6$.

\begin{figure}[t]
\includegraphics[scale=0.3,angle=-90]{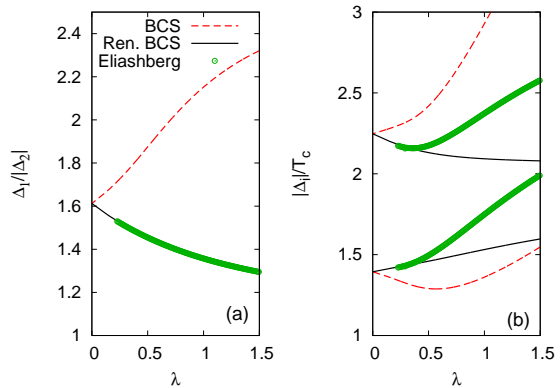}
\caption{(Color online) Evolution of the ratio $\D_1/|\D_2|$ (panel a) and
of $|\Delta_i|/T_c$ (panel b) as function of $\lambda$ calculated by
solving numerically the BCS Eqs. \pref{d1bcs}-\pref{d2bcs}, the
renormalized BCS Eqs.\ \pref{d1bcsren}-\pref{d2bcsren} and the Eliashberg
Eqs. \pref{d1el}-\pref{z1el}, with $B=2.6$.  Notice that while the
renormalized BCS model reproduces the behavior of $\D_1/|\D_2|$ in
Eliashberg theory at all coupling values, the same is not true for the
$|\Delta_i|/T_c$ values.}
\label{fig_gap_vs_l}
\end{figure}

Let us focus first on the gap anisotropy $A\equiv \D_1/|\D_2|$ at $T=0$ as
a function of the average coupling $\lambda\equiv
\sqrt{\lambda_{12}\lambda_{21}}$, which was extensively analyzed in Ref.\
[\onlinecite{dolgov}]. As one can see in Fig.\ \ref{fig_gap_vs_l}a, $A\ra
\sqrt{B}$ as $\lambda\ra 0$,\cite{dolgov} but within the Eliashberg framework
$A$ approaches 1 as $\lambda$ increases, showing that the gaps get closer to
each other. This result is in sharp contrast with the BCS solution, that is
obtained from Eqs. (\ref{d1el})-(\ref{z1el}) by neglecting the equation for
the renormalization functions [$Z_i(n)=1$], and assuming a BCS factorized
square-well model
$D(n-m)=\theta(\omega_0-|\o_n|)\theta(\omega_0-|\o_m|)$. Within this
framework $\Delta_i(n)=\Delta_i\theta(\omega_0-|\o_n|)$ and one gets the
simple equations
\bea
\lb{d1bcs}
\D_1&=&-\lambda_{12}\D_2\Pi_2,\\
\lb{d2bcs}
\D_2&=&-\lambda_{21}\D_1 \Pi_1,
\eea
where $\Pi_i=\pi T \sum_n \theta(\omega_0-|\o_n|)
/\sqrt{\omega_n^2+\Delta_i^2}$. The behavior of $A$ obtained by 
the numerical solution of the previous BCS
set of equations is also reported in Fig.\ \ref{fig_gap_vs_l}a: as one can
see, the two gap values diverge one from the other as the coupling
increases, in contrast to the results of the intermediate-strong coupling
Eliashberg solutions of Eqs.\ \pref{d1el}-\pref{z1el}.\cite{dolgov}

Since the Eliashberg theory accounts for the effects of the
$Z$-renormalization functions, in Ref.\ [\onlinecite{dolgov}] it was
proposed a simple analytical way to illustrate the difference between BCS and
Eliashberg approach by means of a ``renormalized BCS model''.  In this case
the square-well model for the gap equations can be completed with a
corresponding square-well model for the renormalization spectral functions,
$Z_1(n)=1+\lambda_{12}$, $Z_2(n)=1+\lambda_{21}$ for $|\o_n|\le \omega_0$
and $Z_1(n)=Z_2(n)=1$ for $|\o_n|\ge \omega_0$, so that
\bea
\lb{d1bcsren}
\D_1Z_1&=&-\lambda_{12}\D_2 \Pi_2,\\
Z_1&=&1+\lambda_{12},\\
\D_2Z_2&=&-\lambda_{21}\D_1\Pi_1,\\
\lb{d2bcsren}
Z_2&=&1+\lambda_{21}.
\eea
The gap anisotropy obtained from the above set of equations is also shown
in Fig.\ \ref{fig_gap_vs_l}a, compared to the Eliashberg and BCS
solutions. Remarkably, one sees that the renormalization effects account
very well for the decreases of the gap anisotropy $A$ as the coupling
constant increases, giving essentially the same $A(\lambda)$ dependence as
the Eliashberg calculations. As it was suggested in Ref.\
[\onlinecite{dolgov}], this result can be understood by an analytical
approximation of equations \pref{d1bcsren}-\pref{d2bcsren} at low
coupling. Indeed, at $T=0$ one can use the approximate BCS forms of the
$\Pi_i$ bubbles, $\Pi_i=\sinh^{-1}\left({\omega_0}/{|\D_i|}\right)\approx
\log \left({2\omega_0}/{|\D_i|}\right)$ to write a self-consistent
expression for the ratio $A$ as a function of the dimensionless couplings
$\lambda_{12}$ and $\lambda_{21}$:
\be
\lb{eqa}
\frac{AZ_1}{\lambda_{12}}-\frac{Z_2}{A\lambda_{21}}=\log A.
\ee
As a consequence, Eq.\ \pref{eqa} can be solved perturbatively in
powers of the effective coupling
$\tilde{\lambda}=\lambda/\sqrt{Z_1Z_2}$, which takes into account,
at a BCS level, the self-energy renormalization:
\be
\lb{eqexp}
A=\sqrt{\tilde B}(1+c\tilde \lambda+d\tilde \lambda^2)+\OO(\tilde \lambda^3),
\ee
where $\tilde B=(Z_2/Z_1)B$.
%%
%\be 
%\lb{lren}
%\tilde B=\frac{Z_2}{Z_1}B=
%\frac{Z_2}{Z_1}\frac{N_2}{N_1},
%\ee
%
By substituting Eq.\ \pref{eqexp} into Eq.\ \pref{eqa}, and recalling
that $Z_1\sqrt{\tilde B}/\lambda_{12}=Z_2/\sqrt{\tilde
  B}\lambda_{21}=1/\tilde \lambda$,
we immediately obtain:
\be
\lb{coeff}
c=\frac{1}{4}\log \tilde B, \quad
d=\frac{c+c^2}{2}=\frac{4\log \tilde B+\log^2 \tilde B}{32}.
\ee
While making an expansion in the coupling constant one must properly expand
also $\tilde \lambda$ and $\tilde B$ at the leading order in $\lambda$, namely,
\bea
\tilde B&\simeq& B(1+\lambda_{21}-\lambda_{12}),
%\quad \l_\pm=\l_{21}\pm\l_{12}
\\
\tilde\lambda&\simeq&\lambda\left(1-\frac{\lambda_{21}+\lambda_{12}}{2}\right).
\eea
As a consequence, 
one finds\cite{dolgov} for the gaps ratio \pref{eqexp}
\bea
\lb{dolgov}
A=\frac{\D_1}{|\D_2|}
&=&\sqrt{B}\left(1+\frac{\lambda_-}{2}+\frac{\lambda}{4} \log B\right),
\eea
where $\lambda_\pm=\lambda_{21}\pm\lambda_{12}$. The BCS case is recovered
from the previous equations by setting $Z_1=Z_2=1$,
i.e. $\lambda_\pm=0$. Thus, since $\lambda_-<0$ when $B>1$, and the second term
is larger than the third one regardless the value of $B>1$, from Eq.\
\pref{dolgov} it follows that in the BCS case $A$ increases with increasing
$\lambda$, while in the presence of renormalization effects $A$ decreases,
i.e. the two gap values approach each other, as confirmed by the numerical
solutions at all $\lambda$ values reported in Fig.\ \ref{fig_gap_vs_l}a.

From the above considerations and the results of Fig.\ \ref{fig_gap_vs_l}a
one could then be tempted to conclude, as it was done in Ref.\
[\onlinecite{dolgov}], that the McMillan-Allen-Dynes-like equations
\pref{d1bcsren}-\pref{d2bcsren} capture the basic physics of the Eliashberg
solution at all coupling values. However, this is not the case, as we show
in Fig.\ \ref{fig_gap_vs_l}b, where we report explicitly the $\D_i/T_c$
values in the various approaches. Here we use for simplicity in the BCS and
renormalized-BCS case the estimate
$T_c=1.13\omega_0\exp\left(-{1}/{{\lambda}}\right)$ and
$T_c=1.13\omega_0\exp\left(-{1}/{\tilde{\lambda}}\right)$, respectively,
valid at weak coupling by means of the approximate BCS form $\Pi_i=\log
\left({1.13 \omega_0}/{T_c}\right)$ of the bubbles near $T_c$.  As one can
see in Fig.\ \ref{fig_gap_vs_l}, in the renormalized BCS case the gap
values approach each other by a decrease of the larger $\D_1/T_c$ value,
and a partial increase of the smaller $|\D_2|/T_c$ value.  This result can
be again understood analytically at low coupling by resorting to the above
expansion \pref{eqexp} and the $T_c$ expression. We then obtain the leading
dependence of $\D_i/T_c$ on the coupling:
\bea 
\frac{\D_1}{T_c}&=&1.76\tilde B^{1/4}\left(1+\tilde\lambda
\frac{4\log\tilde B-\log^2\tilde B}{32}\right) \nn\\ 
\lb{d1tceli} &=&1.76
B^{1/4}\left(1+\frac{\lambda_-}{4}+\lambda \frac{4\log B-\log^2 B}{32}\right),\\
\frac{|\D_2|}{T_c}&=&1.76\tilde B^{-1/4}\left(1-\tilde\lambda \frac{4\log\tilde
B+\log^2\tilde B}{32}\right)\nn\\ 
\lb{d2tceli} &=&1.76
B^{-1/4}\left(1-\frac{\lambda_-}{4}-\lambda\frac{4\log B+\log^2 B}{32}\right). 
\eea
%
%%%%%%%%%%%%%%%%%%%%%%%%%%%%%%%%%%%%%%%%%%%%%%
%%%%%%%%%%%%%%%%%%%%%%%%%%%%%%%%%%%%%%%%%%%%%%
Also in this case the terms in $\lambda_-<0$ are larger than the others, so
that one recovers from the above equations that at low coupling $\D_1/T_c$
decreases and $|\D_2|/T_c$ increases as a function of $\lambda$. However,
except for a very narrow range of coupling $\lambda \lesssim 0.2-0.3$, the
numerical Eliashberg solution of Eqs.\ \pref{d1el}-\pref{z1el} is markedly
different. Indeed, in the intermediate/strong coupling regime, which is the
one relevant for pnictides, both $\Delta_1/T_c$ and $|\Delta_2|/T_c$ {\em
increase} with $\lambda$ in the Eliashberg case. Thus, in the full
numerical Eliashberg solution the gaps in the two bands approach each-other
by means of an increase of the absolute $\D_i/T_c$ ratio in both the bands,
that is the typical signature of strong coupling. We note in passing that
while Fig.\ \ref{fig_gap_vs_l}a reproduces the findings of Ref.\
[\onlinecite{dolgov}], the same is not true for Fig.\
\ref{fig_gap_vs_l}b. Indeed, the numerical results for $\D_i/T_c$ in the
Eliashberg theory reported in Ref.\ [\onlinecite{dolgov}] differ
significantly from our findings, even in the low-coupling regime where the
above analytical analysis supports completely our numerical calculations.

\section{Superconducting properties in a four-band model}

\begin{figure}[t]
\includegraphics[scale=0.2,angle=0]{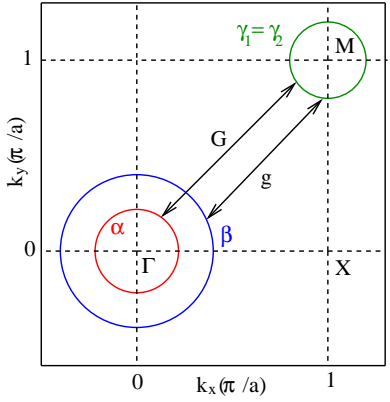}
\caption{(Color online) Schematic view of the four-band model we use for
  pnictides. $\a$ and $\b$ are the hole bands around the $\G$ point,
  $\g_1=\g_2$ are the two degenerate electron bands at the M point. The
  sizes of the pockets are inferred from the experiments in
  Ba$_{0.6}$K$_{0.4}$Fe$_2$As$_2$.\cite{ding2} The
  coupling anisotropy, with $G>g$ is suggested by the different nesting
  properties of electron to the hole bands, as due to their different
  sizes.}
\label{fig_pockets}
\end{figure}

In the previous section we have shown within a simple two-band model that a
full numerical approach is needed to capture the property of Eliashberg
equations of removing the DOS anisotropy of the gaps in the presence of
interband pairing. Following the same reasoning we focus in this section on
a four-band model, to correctly capture the physics of pnictides.  In
particular, we shall discuss the case of hole-doped
Ba$_{0.6}$K$_{0.4}$Fe$_2$As$_2$, using the
notation of Refs.\ [\onlinecite{ding1,ding2}], where $\a$ ($\b$) is the inner
(outer) hole-pocket centered around the $\Gamma$ point, and
$\g_1,\g_2$ are
the two electron-like pockets centered around the M points of the folded
Brillouin zone of the FeAs planes (see Fig.\ \ref{fig_pockets}).
The dominant interactions in pnictides are thought to be mainly interband
ones, connecting hole Fermi-sheets with electron Fermi-sheets, through the
exchange of spin fluctuations at the antiferromagnetic wave-vector ${\bf
Q}=(\pi,\pi)$.\cite{arita,chubukov} The strength of such interband coupling
between hole and electron pockets, in these materials, is in addition
expected to depend on the relative sizes of the pockets, that naturally
affect the nesting condition for the spin fluctuations mediating the
pairing.  In particular, in BKFA the size of the $\a$ and $\g_i$ Fermi
surfaces is quite comparable, while the $\b$ band has a Fermi surface
substantially larger, with a corresponding less degree of nesting.  In this
situation the interband $\b-\g_i$ coupling $g$ is expected to be
significantly smaller than the $\a-\g_i$ coupling $G$.  In addition, since
the two electron pockets have comparable sizes, we assume them for
simplicity to be degenerate, and we denote with $N_\g=N_{\g_1}+N_{\g_2}$
the total DOS in the electron pockets.  We can write thus the multiband
Eliashberg equations in the form:
\bea 
\lb{dael}
Z_\a(n)\Delta_\a(n)&=& -\pi T \sum_m D(n-m) \frac{GN_\g
\Delta_\g(m)}{\sqrt{\omega_m^2+\Delta_\g^2(m)}},\\
\lb{dbel} 
Z_\b(n)\Delta_\b(n)&=& -\pi T \sum_m D(n-m)
\frac{gN_\g
\Delta_\g(m)}{\sqrt{\omega_m^2+\Delta_\g^2(m)}},\\
\lb{dgel} 
Z_\g(n)\Delta_\g(n)&=& -\pi T \sum_m D(n-m)
\left[
\frac{GN_\a \Delta_\a(m)}{\sqrt{\omega_m^2+\Delta_\a^2(m)}}\right.
\nn\\ 
&&+\left.
\frac{gN_\b \Delta_\b(m)}{\sqrt{\omega_m^2+\Delta_\b^2(m)}}\right],
\eea 
\bea \lb{zael}
Z_\a(n)&=& 1+\frac{\pi T}{\omega_n} \sum_m
D(n-m) \frac{GN_\g\omega_m}{\sqrt{\omega_m^2+\Delta_\g^2(m)}},
\lb{zbel} 
\\
Z_\b(n)&=& 1+\frac{\pi T}{\omega_n} \sum_m D(n-m)
\frac{gN_\g\omega_m}{\sqrt{\omega_m^2+\Delta_\g^2(m)}},\\ 
\lb{zgel}
Z_\g(n)&=& 1+\frac{\pi T}{\omega_n} \sum_m D(n-m)
\left[
\frac{GN_\a\omega_m}{\sqrt{\omega_m^2+\Delta_\a^2(m)}}\right .\nn\\
&&\left.
+\frac{gN_\b\omega_m}{\sqrt{\omega_m^2+\Delta_\b^2(m)}}\right] ,
\eea
In the BCS limit, the above equations reduce to the one already discussed
in Ref.\ [\onlinecite{benfatto_4bands}]:
\bea
\D_\a&=&-N_\g G\D_\g \Pi_\g\\
\D_\b&=&-N_\g g\D_\g \Pi_\g\\
\D_\g&=&GN_\a\D_\a \Pi_\a+g N_\b\D_\b \Pi_\b,
\eea
and $T_c$ is given by $T_c=1.13\omega_0e^{-1/\Lambda}$, where,
in analogy with the two-band case,
we introduce the effective coupling 
\be
\lb{leff}
\Lambda=\sqrt{G^2N_\g N_\a+g^2N_\g N_\b}.
\ee
It is interesting to notice that in the BCS limit the ratio between the
two hole gaps is simply given by the ratio of the corresponding interband
couplings, i.e. $\D_\a(n)/\D_\b(n)=g/G$, {\em independently} on the
relative DOS. The simple experimental observation $\D_\a/\D_\b\approx 1/2$
would suggest thus, in BCS, $g \approx G/2$.  Such constraint does not
apply however to a more accurate Eliashberg analysis
[Eqs. (\ref{dael})-\pref{zgel}], where
$\D_\a(n)/\D_\b(n)=(g/G)Z_\b(n)/Z_\a(n)$, so that, in
principle, the ratio of the gaps in the two hole pockets depends both on
the couplings and on the DOS of the various bands, requiring thus a more
careful analysis.

We employ now Eqs. (\ref{dael})-(\ref{zgel}) to evaluate the
microscopic interband couplings $G$, $g$ from the physical constraints
given by the experimental determination of the gap magnitudes on the
different bands. We shall apply later this analysis to calculate
different superconducting and normal-state properties, as the superfluid
density and the specific heat, in order to have an independent check of the
reliability of our analysis.

A debated issue in this context is the assess of a proper choice for the
underlying normal state electronic bands.  Indeed, as we mentioned in the
introduction, ARPES measurements in several pnictide families report
significant differences in the electronic dispersion compared with LDA
calculations, with an apparent renormalization of the whole band structure
by a factor 2.\cite{ding2,lu,yang,yi} Most striking, such band narrowing
seems to be operative up to very high energy scales, as it is confirmed
also by recent optical sum-rule analysis performed in LaFePO
samples.\cite{basov_sumrule_09} This overall renormalization of the bands
with respect to LDA seems thus a general feature of pnictides, probably
arising from local Hubbard-like correlations,\cite{georges,vollhardt,yang2}
so that it cannot be captured by the coupling of the electrons to
low-energy bosonic modes.  To take into account this feature, we estimate
our input band parameters directly from ARPES experiments, that have enough
resolution to capture the high-energy mass renormalization. Using the
tight-binding fit of the bands reported in Ref.\ [\onlinecite{ding2}], we
approximate close to the Fermi level each band as
$\e_i(\bk)=\e_i^0-t_i|\bk|^2$, where $\bk$ is measured with respect to the
$\G$ point for the hole bands and to the $M$ point for the electron bands
(see Fig.\ \ref{fig_pockets}). $N_i=1/4\pi t_i$ is the DOS (per spin) in
each band. The band parameters for each band, extracted from Ref.\
[\onlinecite{ding2}], are listed in Table \ref{t-table}.
\begin{table}[t]
\begin{center}
\begin{tabular}{cccc}
\hline \hline
& $\a$& $\b$ & $\g_1,\g_2$\\
\hline
$\e_i^0$ (meV) & 28 & 43 & $-60$ \\
\hline
$t_i$ (meV) & 54 & 27.5& 160 \\
\hline
$N_i$ (eV$^{-1}$)& 1.47 & 2.89 & 0.50\\
\hline
$n_i$& 0.082 & 0.24 & 0.06\\
\hline
\hline
\end{tabular}
\end{center}
\caption{Microscopic band parameters extracted from
  Ref.\ [\onlinecite{ding2}] by approximating each band with a parabolic
  form $\e_i(\bk)=\e_i^0-t_i |\bk|^2$. $N_i=1/4\pi t_i$ is the DOS and
  $n_i$ is the number of holes/electrons per unit cell in each band. Note
  that for the $\g_1,\g_2$ bands we accounted for the corrections due to
  the elliptical shape reported in Ref.\ [\onlinecite{ding2}].}
\label{t-table}
\end{table}
To model the spin-mediate interaction, following Ref.\ [\onlinecite{millis}]
and the recent report [\onlinecite{bourges_spin}] we use
$B(\Omega)=\Omega\omega_0/\pi(\omega_0^2+\Omega^2)$, with the
characteristic energy scale $\omega_0=20$ meV estimated from experimental
measurements.\cite{matan,osborn,christianson,bourges_spin} 

\begin{figure}[t]
\includegraphics[scale=0.3,angle=0]{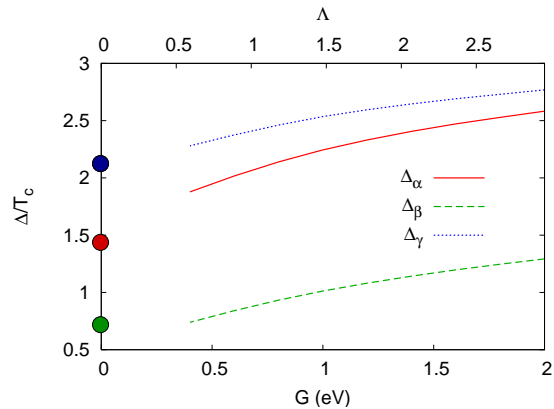}
\caption{(Color online) Dependence of the ratios
  $\D_\a/T_c$, $\D_\b/T_c$ and $\D_\g/T_c$ as functions of the coupling $G$
($g=G/2$)
within the Eliashberg theory. The
  dots at $G=0$ mark the weak-coupling BCS values.}
\label{fig_4gap_vsl}
\end{figure}

In approaching a numerical analysis, we should note that, neglecting for
the moment the weaker coupling $\g-\b$, the model is equivalent to two-band
case discussed in the previous Section, with $B=N_\a/N_\g\simeq 1.5$, so
that at the BCS level the gap values in the two strongly-nested bands are
different at least by $20\%$.  The presence of a finite scattering $\b
\leftrightarrow \g$ makes this scenario even more complex, with the onset
of the gap $\Delta_\b$, which contributes to increase the magnitude of
$\Delta_\g$. In the absence of any particular degeneracy between the $\a$
and $\g$ bands, this situation would result, within a weak-coupling BCS
framework, in the appearance of three different gaps on the different Fermi
sheets, as marked by the dots in Fig.\ \ref{fig_4gap_vsl}.  The
experimental observation from ARPES of two nearly degenerate gaps on the
$\a$ and $\g$ bands calls then for further investigation.

In this regard, the strong coupling results from the two-band model,
discussed in Ref.\ [\onlinecite{dolgov}] and readdressed more specifically in
Section II, shed a new interesting light once plugged in a four-band
model.  Within this context, indeed, the nesting-driven strong coupling
interaction between the $\a$ and $\g$ bands leads to a merging of the value
of the two large gaps $\Delta_\a$, $\Delta_\g$ in the systems. At the same
time, the {\em anisotropy} of the coupling between the two hole pockets and
the electron one, guaranteed by the different nesting conditions, allows
the system to keep the gap in the $\beta$ band smaller even within the
Eliashberg approach, in agreement with ARPES.

To have a more quantitative insight we plot in Fig. \ref{fig_4gap_vsl} the
gap values $\Delta_i=\Delta_i(n=0)$ obtained from the numerical solution at
$T=T_c/20$ of Eqs.\ \pref{dael}-\pref{zgel} as a function of the coupling
$G$, for the indicative case $g=G/2$, which would gives the experimental
value $\D_\b=0.5 \D_\a$ in the BCS limit $G \rightarrow0$.  As we can see,
the use of a four-band model is crucial to recover the hierarchy of the
gaps: while in the two-band case by increasing $\lambda$ one is forced to
have a single gap value, in the four-band model the anisotropy of the
couplings, which follows naturally from the different nesting properties of
the various bands, allows for a two-gap result, where two almost degenerate
larger gaps are predicted on the electron and on the inner hole pocket,
while a smaller gap is found in the outer hole pocket.

As mentioned in the introduction, whereas the experimental gaps in the $\a$
and $\g$ bands are approximately equal, the ratio $\D_\a/T_c\approx 3.75$
is quite larger than the BCS value 1.76. Recently, the possibility to
reproduce exactly these ratios in BKFA has been investigated in Ref.\
[\onlinecite{umma}], where it was shown that an extremely large value of
the effective coupling $\Lambda\gtrsim 4$ and a small boson energy scale
$\omega_0=10$ meV were required.  This kind of analysis needs however to be
taken with some caution. Indeed, any estimate of $T_c$ in a mean-field-like
theory as the Eliashberg one is in general questionable in two-dimensional
systems, as pnictides, where the superconducting fluctuations, when
properly taken into account, could significantly reduce
$T_c$.\cite{varlamov} For this reason, while the low-temperature gap values
estimated within a mean-field-like Eliashberg theory can be quantitatively
sound, the estimate of $T_c$ done within the same approach must be taken as
an upper limit. Moreover, the values of $\omega_0$ and of the coupling do
not influence only the ratios $\D_i/T_c$, but they also control in a
crucial way other physical quantities, like the mass renormalization and
the position of the kink in the band dispersion, that can be accessed
experimentally. For these reasons we investigate here the following
approach: we discard the exact determination of $T_c$ and we choose the
coupling strength as to reproduce the magnitude of the different gap
values. Afterwards, we check if by means of the same parameter
values we can account for other experimental results related to
mass-renormalization effect in the spectral and thermodynamic properties.
\begin{table}[t]
\begin{center}
\begin{tabular}{cccc}
\hline \hline
& $\a$& $\b$ & $\g_1,\g_2$\\
\hline
$\D_i$ (meV) & 9.48 & 4.35 & $-10.48$ \\
\hline
$Z_i$ (meV) & 2.09 & 1.35& 3.67 \\
\hline
$m_i^*/m_e$& 9.61 & 12.28 & 5.72\\
\hline
$J_{s,i}(0)$ (meV)& 4.8 & 10.7 & 6.14\\
\hline
$\D^{\rm exp}_i$ (meV)& $12\pm 1$ & $6\pm 1$ & $12 \pm 1$\\
\hline
\hline
\end{tabular}
\end{center}
\caption{Eliashberg parameters obtained by numerical solutions of the
  Eqs.\ \pref{dael}-\pref{zgel} with the coupling matrix \pref{couplings}.
  In the last row we report for comparison also the experimental values of
  the gaps from Ref.\ [\onlinecite{ding1,ding2}].}
\label{t-js}
\end{table}

In Table \ref{t-js} we summarize our results for the interband scattering,
$G=1.1$ eV , $g=0.32 G\simeq 0.35$ eV and $\omega_0=20$ meV,
obtained to reproduce the experimental gaps.
The multiband matrix of the coupling constants,
in the band space $(\alpha,\beta,\gamma_1,\gamma_2)$ reads:
\be
\lb{couplings}
\hat{\lambda}=
\left(
\begin{array}{cccc}
0 & 0 & 0.55 & 0.55\\
0 & 0 & 0.18 & 0.18\\
1.62 & 1.01 & 0 & 0\\
1.62 & 1.01 & 0 & 0
\end{array}\right),
\ee
and the effective average coupling defined in Eq.\ \pref{leff} is
$\Lambda=1.5$.  We notice that a better agreement with the experimental gap
values could be enforced by slightly different DOS than the ones estimated
in Ref.\ [\onlinecite{ding2}], or by assuming a non-zero interband
coupling, as due to phonons. However, we prefer here to use a minimal set
of free parameters to show the overall quantitative agreement between our
approach and the experiments.  We obtain a critical temperature
$T_c=0.21\omega_0=48.7$ K, which overestimates the experimental one
$T_c^{\rm exp}=37$ K by about 10 K.  Taking into account that a recent
analysis of paraconductivity has shown that this is the typical range of
temperature where superconducting fluctuations are active in
pnictides,\cite{benfatto_fluct} one can expect that the effect of
superconducting fluctuations beyond Eliashberg theory will contribute to
improve the agreement between the present estimate of $T_c$ and the
experimental value.

%

%

%\begin{figure}[htb]
%\includegraphics[scale=0.3,angle=-90]{bands_ren.eps}
%\caption{(Color online) Renormalized bands within Eliashberg theory at
%  $T>T_c$, The kink in the band dispersion, that is due to the mass
%  renormalization at low energy, occurs at $\e=\omega_0$, as indicated by
%  the arrows.}
%\label{fig_bands}
%\end{figure}
%

The coupling of carriers to spin fluctuations reflects on the one-particle
spectral properties already above $T_c$.  In particular, the bands in the
normal state are expected to display a kink at an energy $\omega_0$, so
that for $E<\omega_0$ the Fermi velocity is renormalized with respect to
the bare value, with $v_{{\rm F},i}^*=v_{{\rm F},i}/Z_i$ (or equivalently
$m_i^*=Z_i m_{{\rm b},i}$), where $Z_i=Z_i(n=0)$ and $m_{{\rm b},i}$ is the
band mass of the sheet $i$.
\begin{figure}[t]
\includegraphics[scale=0.35,angle=-90]{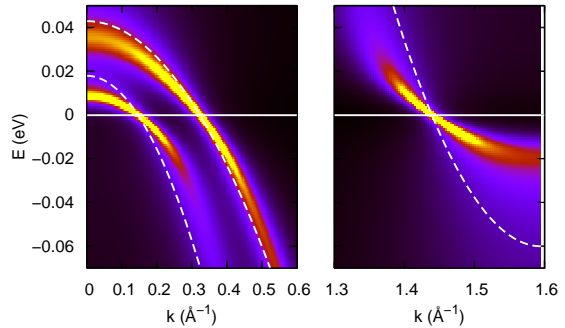}
\caption{(Color online) Intensity map of the spectral function in the
  normal state for the
  renormalized bands within the Eliashberg theory, as compared to the bare
  band dispersions, represented by the dashed lines. Note that
  within a spin-fluctuation model for the bosonic excitation the kink
  features at $E=-\omega_0$ are considerably smeared out in the spectral
  functions. Here we added a constant broadening $\eta=0.002\omega_0$
  to account for a small amount of disorder.}
\label{fig_spectra}
\end{figure}
In Fig. \ref{fig_spectra} we show the intensity map of the spectral
function for the interacting hole and electron bands, as obtained by the
Marsiglio-Schossmann-Carbotte analytical-continuation procedure,\cite{msc}
along with the bare band dispersions. For the spin-fluctuation model used
here, and for the coupling values deduced by the measured gaps, the kinks
at $\omega_0$ are significantly smeared out in the spectral function.
Unfortunately, the experimental resolution of the data in Ref.\
[\onlinecite{ding2}] is not high enough to resolve the effect of low-energy
spin fluctuations from the high-energy renormalization.  However, it is
interesting to notice that a similar kink has been actually observed in
high-resolution ARPES measurements performed by an other group in a BFKA
sample with lower doping than the one we are discussing
here.\cite{borisenko} In particular, the authors find a kink around
approximately 30 meV, and a velocity renormalization in the inner and outer
hole pockets of about 1.8 and 1.6. Thus, given the difference in doping and
the lack in our approach of a specific momentum dependence of the
spin-exchanged fluctuations (that can contribute to
slightly increase the effective energy of the kink), our results are in good
agreement with these findings.
%We notice that due to the relatively large value of the effective
%coupling and the small value of the Fermi energy the effect of the kink is
%visible in the bands up to the top/bottom for the hole/electron bands
%respectively, see Fig.\ \ref{fig_bands}. 
%As a consequence, a value of the
%coupling as large as the one suggested in Ref.\ \cite{umma} would lead to a
%complete depressions of the spectra near the band edge, that could not be m
%that is not seen in the experiments. 

Below the superconducting transition the position of the kink in the band
dispersion is in principle shifted in each band to an energy given by
$\omega_0+\D_j$, where $\D_j$ is the value of the gap in the bands coupled
to the $i$-th sheet.  In particular, in the $\a,\b$ bands the kink due to
spin fluctuations is expected to be recovered at an energy
$\o_0+\D_\g\approx 30$ meV, while in the $\g_1,\g_2$ bands two kinks will
appear, at the energies $\o_0+\D_a\approx 30$ meV and $\o_0+\D_b\approx 25$
meV.  It should be pointed out, however, that, because of the smearing of
the spectra in Fig.\ \ref{fig_spectra}, it would be quite hard to detect
the possible splitting in the $\g$ bands due to two different gap values of
the $\a$ and $\b$ bands.  At the best of our knowledge, a clear
identification below $T_c$ of a continuous shift at higher energy of the
kinks observed in the normal state has not been reported yet in
pnictides.\cite{note_kink} 

Finally, it is worth mentioning that an additional source of discrepancy
between the LDA and the experimental bands in pnictides comes from
finite-band effects, that have been discussed in Ref.\
[\onlinecite{cappelluti_dhva}].  Indeed, when the strong particle-hole
asymmetry of pnictides is taken into account by considering the finite
bandwidth, the spin-mediated interband coupling leads to a shift of the
Fermi momenta with respect to LDA, that has been indeed measured in other
pnictides by de Haas-van Alphen experiments.\cite{coldea,analytis} In the
present case we did not compute explicitly these shifts, since they are
already included in the band dispersion extracted from the
measured ARPES data.  Since finite-band effects do not alter qualitatively
the self-energy corrections apart from the mentioned energy shift, we
solved Eqs.\ \pref{dael}-\pref{zgel} within the usual infinite-band
approximation.

\section{Thermodynamic properties}

Having determined an appropriate set of band parameters and multiband
couplings, we investigate now the effects of the spin-mediated interactions
on the thermodynamical properties. Indeed, the signatures of low-energy
renormalization are much more easily detectable in thermodynamic
measurements of masses enhancement than in photoemission, where a very high
resolution is required to resolve the kinks in the band dispersion.

Let us consider as a first insight specific-heat measurements.  In the
normal state, the coefficient $\g_N=C_V/T$ of the linear $T$ term in the
specific heat measures the mass enhancement, once
compared with the value estimated for non-interacting bands.  To clarify
the units, we shall refer in the following to the specific heat per formula
unit (so that one mole of the materials contains 2 Fe atoms). By expressing
the renormalized DOS (per spin) of each band as a function of the
renormalized mass $m^*$, $N_i^*=m_i^*/2\pi$ and, restoring
explicitly all the needed dimensional constants, we
have that each band contributes to $\g_N$ as:
\be
\lb{defg}
\g_{N,i}=\frac{2\pi^2 k_B^2}{3}N_i^*N_A a^2=
1.5 \frac{m_i^*}{m_e} \, \mathrm{mJ/K^2
mol},
\ee
where $a=3.9$ \AA \ is the lattice spacing in BKFA, $N_A$ is the Avogadro
number, $k_B$ the Boltzmann constant and $m_e$ the free electron mass.
Within LDA one obtains $\g_N=9.26$ mJ/K$^2$mol,\cite{ma_lda_bkfa} that is
remarkably smaller than the values obtained in doped BKFA, either by direct
analysis of the normal-state specific heat $\g_N\approx 49$
mJ/K$^2$mol\cite{loid_cm09} or by measurements of the upper critical field
$\g_N\approx 63$ mJ/K$^2$mol\cite{mu}.  By means of the band parameters
extracted from ARPES and listed in Table I one estimates $\g_N=25$
mJ/K$^2$mol, while with the renormalized masses listed in Table II, that
include low-energy renormalization effects on the ARPES bands, we can
estimate $\g_N=50$ mJ/K$^2$mol, in very good agreement with Ref.\
[\onlinecite{loid_cm09}]. Thus, the additional mass renormalization due to
the spin-fluctuations exchange is fundamental to reconcile ARPES and
specific-heat measurements. We note that the effective masses obtained
until now in the 1111 family by means of de Haas-van Alphen experiments are
significantly smaller than the values reported in Table I.\cite{coldea}
This can be attributed to smaller coupling strength, consistently with the
smaller $T_c$ values of the 1111 pnictides (see also Ref.\
\onlinecite{cappelluti_dhva}). Recently, de Haas van Alphen experiments in
BaFe$_2$As$_{1-x}$P$_x$ show a tendency of considerably increase of the
mass enhancement for this 122 compound close to the optimal $T_c\simeq 30$ K
value.\cite{carrington_cm09} Further de Haas-van Alphen experiments are
required to establish the correlation between the mass enhancement and the
$T_c$ values that our analysis suggests.

To complete the analysis of the specific heat, we compute  numerically the
free energy difference $\Delta F_i$ per band between the superconducting
and the normal state according to the expression given in Ref.\
[\onlinecite{carbotte}], namely
\begin{eqnarray}
\Delta F_i(T)
&=&
-\pi T \sum_{n} N_i
\left[
Z^{\rm S}_i(n)-\frac{Z^{\rm N}_i(n)|\omega_n|}
{\sqrt{\omega_n^2+\Delta_i^2(n)}}
\right]
\nonumber\\
&&\times
\left[\sqrt{\omega_n^2+\Delta_i^2(n)}-|\omega_n|\right],
\label{free}
\end{eqnarray}
where $Z^{\rm S}_i$, $Z^{\rm N}_i$ are the $Z$-renormalization functions
for the $i$-band
calculated in the superconducting and in the normal state
respectively,
and we evaluate the specific-heat difference as
$\Delta C_{V,i} = -T\partial^2 \Delta F_i/\partial T^2$.
\begin{figure}[t]
\includegraphics[scale=0.3,angle=-90]{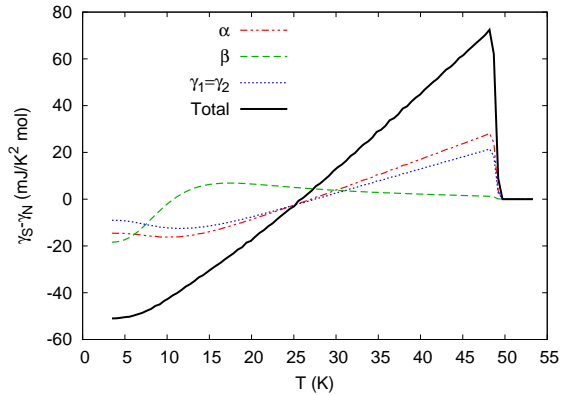}
\caption{(Color online) Temperature dependence of the difference between
  the superconducting and normal-state specific-heat coefficient 
$\g_S-\g_N=\Delta C_V/T$ in the
  various bands, along with the total one.}
\label{fig_cv}
\end{figure}
In Fig.\ \ref{fig_cv} we report the temperature dependence of the ratio of
the specific heat to temperature for each band, along with the total one
$\g_S-\g_N=\Delta C_V/T=(\sum_i\Delta C_{V,i})/T$.  It should be stressed
that, given the interband nature of the scattering, the decomposition of
the total specific heat $\Delta C_V$ in single band contributions $\Delta
C_{V,i}$ is purely formal since all bands are coupled and the free-energy
of one band depends implicitly on the properties of all the other ones.  In
agreement with what expected in multiband superconductors (see for example
Ref.\ [\onlinecite{bouquet,kogan}]), at low temperature the increase of
$\g_S$ is controlled by the quasiparticle excitations across the smaller
gap, i.e. $\D_\b$ in our case, while at higher temperature also
quasiparticles excitations in the bands with larger gap are thermally
activated. The presence of multiple energy scales for the quasiparticle
excitations can lead to the presence of humps in the intermediate
temperature range between $T=0$ and $T_c$, as observed for example in
MgB$_2$ compounds.\cite{bouquet,kogan} In our case, such hump is most
evident in the contribution of the band $\beta$ which is related to the
smaller gap. Such hump is not however clearly reflected in the total
specific heat since it is partially compensated by the depletion in the
other band contributions.  The physical origin of such depletions is
however questionable since it would give rise to a negative contributions
to the specific heat for some band, and it has been argued that it is
related to the neglecting of superconducting effects in the boson
propagator itself, as a consequence of the electron-boson renormalization.
\cite{mansor} The restoring of a physical positive contribution of the
specific heat for all the bands could then make the hump of the
$\beta$ band evident also in the total specific heat. It is worth noting
that it is not clear yet if such a hump is visible in the experiments,
since it is observed in Ref.\ [\onlinecite{mu}] but not in Ref.\
[\onlinecite{loid_cm09}],
where a temperature profile remarkably similar to the calculations shown in
Fig.\ \ref{fig_cv}  is reported. Thus, more theoretical and experimental
work is required to establish the real temperature profile of the specific
heat in pnictides. 
Finally, we estimate the jump of the specific heat
at the transition as $\D C_V/T_c=72.5$ mJ/K$^2$mol, so that $\D C_V/\g
T_c=1.45$. Both are in good agreement with the experimental estimates of $\D
C_V/T_c=98$ mJ/K$^2$mol and $\D C_V/\g T_c=1.5$.\cite{mu} As far as $\D
C_V/\g T_c$ is concerned, it must be noticed that even though this estimate
is apparently near to the {\em single-band} BCS value 1.43, actually 
in the BCS {\em multiband} case the ratio $\D C_V/\g T_c$ is no more
universal, so that the experimental result cannot be taken as indicative
that pnictides are in the weak-coupling regime.

The effect of the multiple gaps is present also in the temperature
dependence of the superfluid density $n_s$, which is experimentally accessible
from measurements of the penetration depth $\lambda_L$, through the relation
$\lambda_L^{-2}=4\pi e^2 n_s/m c^2$. In two dimensional systems one
defines conventionally an energy scale $J_s$ associated to the superfluid
density $n_s^{2d}\equiv n_s d$ of each plane such that:
\be
\lb{js}
J_s=\frac{\hbar^2 n_s^{2d}}{m }= 
\frac{\hbar^2 c^2 }{4\pi  e^2 }\frac{d}{\lambda_L^2}\Ra
J_s[\mathrm{K}]=16.37\frac{1}
     {\lambda^2_L [(\m \mathrm{m})^2]}\, 
\ee
where $d=6.6$ \AA \ is the interlayer spacing. Within the Eliashberg approach
and in the clean case,
$J_s$ can be computed in each band as:\cite{carbotte}
\be
\lb{rhos}
J^i_s(T)=2 N_i (v_{{\rm F},i} a)^2 \pi k_B
T\sum_{n=1}^\infty \frac{\D_i^2(n)}
{Z_i(n)  [\o_n^2+\D_i^2(n)]^{3/2}} .
\ee
Since we are using a parabolic approximation, at $T=0$ the superfluid
density coincides with the carrier density $n_i$ in each band, 
so that Eq.\ \pref{rhos} reduces to the standard formula 
\be
\lb{rhos0}
J^i_s(T=0)=\frac{\hbar^2 n_i}{m_i^*}=\frac{2t_in_i}{Z_i}.
\ee
For the band parameters listed in Tables \ref{t-table}-\ref{t-js},
we can then estimate at
$T=0$ $J_s=\sum_i J_s^i=301$ K, while the unrenormalized value (obtained
with the ARPES band values but without taking into account the
spin-fluctuations induced mass enhancement) would be $J_s\simeq 700$ K. 

\begin{figure}[t]
\includegraphics[scale=0.3,angle=-90]{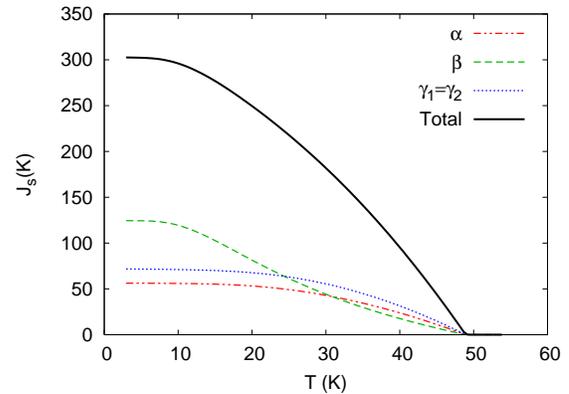}
\caption{(Color online) Temperature dependence of the superfluid density in
  the four bands within Eliashberg theory.}
\label{fig_rhos}
\end{figure}

The temperature dependence of $J_s$ is
reported in Fig.\ \ref{fig_rhos}, 
and it does not differ considerably from
the profile obtained within a simpler multiband BCS
approach.\cite{benfatto_4bands,choi} In all the bands the $J^i_s(T)$ have a
flat temperature dependence at low $T$, typical of exponentially activated
quasiparticle excitations across the constant superconducting gaps.
In the $\b$ band the
deviations from the single-band case are more pronounced, due to the low
$\D_\b/T_c$ value. This anomaly is reflected also in the total superfluid
response, that differs from the standard single-band case. We notice that
the $\a$ and $\g$ bands have almost the same normalized profile
$J_s(T)/J_s(0)$, due to the fact that the gaps have approximately the same
value in these two bands. Thus, the curve in Fig.\ \ref{fig_rhos} does not
differ qualitatively by a BCS two-band calculation, performed assuming that
the band with the large gap contributes to 63$\%$ of the total superfluid
density. This is the reason why two-band BCS fits (implemented with non-BCS
values of $\D_i/T_c$) work usually quite well in the comparison with the
experimental data.\cite{hiraishi,khasanov}

The comparison of our predictions with the experimental data is a quite
delicate issue due to the presence of many extrinsic effects which can
spoil a robust experimental determination of $J_s(T)$.  On one hand,
indeed, measurements of microwave surface-impedance have direct access only
to $\lambda_L(T)-\lambda_L(0)$,\cite{matsuda,prozorov} so that the
determination of the $\lambda_L^{-2}(T)$ profile usually requires the
separate knowledge of $\lambda_L(0)$, even if in some cases one can
directly access the normalized penetration depth
$\lambda^2_L(0)/\lambda^2_L(T)$.\cite{matsuda} On the other hand, the
$\m$SR measurements are also quite delicate because the signal due to the
screening supercurrents must be disentangled from the signal due to
magnetic domains, an issue particularly delicate in those samples where
superconductivity coexists with residual magnetic
ordering.\cite{uemura_high} A third problem comes from the presence of
disorder, which is particularly severe in pnictides due to the $s_\pm$
symmetry of the order parameter, so that inter-band impurity scattering is
pair-breaking and acts in the same way as magnetic impurities in a
conventional single-band $s$-wave superconductor. As a consequence,
disorder can lead to a change of the low-temperature superfluid-density
profile from the exponential behavior to a power-law behavior in the
strong-impurity limit.\cite{chubukov_rhos,bang} When this is the case, one
observes also a strong suppression of $J_s(0)$ with respect to the
clean-limit estimate based on Eq.\ \pref{rhos0}.\cite{prozorov} To give a
hint about how much spread of the data is present in the literature we
summarize some results for hole-doped BKFA in Tab. \ref{t-refs}.
\begin{table}[t]
\begin{center}
\begin{tabular}{cccccc}
\hline \hline
Ref. & $x$& $T_c$ (K)& $\lambda_L(0)$ ($\mu$m) & $J_s(0)$ (K)& temp. dep.\\
\hline
\onlinecite{hiraishi} & 0.4 & 38 & 0.231 & 307 & exp \\
\hline
\onlinecite{uemura_low} & 0.45 & 30 & 0.569 & 51 & exp \\
\hline
\onlinecite{uemura_high} & 0.5 & 37 & 0.298 & 184 & pow \\
\hline
\onlinecite{li} & 0.4 & 37 & 0.208 & 380 & - \\
\hline
\onlinecite{khasanov} & - & 32 & 0.327 & 153 & exp \\
\hline
\onlinecite{prozorov} & 0.45 & 30 & 0.600& 45 & pow \\
\hline
\hline
\end{tabular}
\end{center}
\caption{Summary of some superfluid density measurements in BKFA compounds
in the literature. The last column indicate the best fit of the temperature
dependence at low $T$, where 'exp' stays for (two-band) exponential fit,
and 'pow' stays for power-law.}
\label{t-refs}
\end{table}
Note that the exact doping of the samples is not always available, and that
for the same nominal doping $T_c$ can be sensibly different, due to the
different level of disorder.  We also included an estimate of
$\lambda_L(0)$ done in Ref.\ [\onlinecite{li}] from optical-conductivity
data. To make a comparison with our clean-limit estimate \pref{rhos0} of
$J_s(0)$ we should then disregard the data characterized by a power-law
behavior at low temperature, signature of strong interband impurity
scattering, and the data taken for samples with considerably different
$T_c$. The best candidates are then the data from Refs.\
[\onlinecite{hiraishi}] and [\onlinecite{li}], which are taken in samples
with the same doping level $x=0.4$ and same $T_c=37$-38 K.  These
measurement give $J_s(0)=307-380$ K, which is in very good agreement with
our Eliashberg calculations.

Finally, we would like to mention that despite the spread of experimental
data, in pnictides the $T=0$ value of the superfluid density is
approximately consistent with the value of the Fermi energy, apart the not
too large renormalization effects discussed here.  As it was emphasized
already in Ref.\ \onlinecite{benfatto_4bands}, this implies generically low
values of the superfluid density in pnictides, simply due to the low value
of the density of carriers in each band, see Eq.\ \pref{rhos0}. This result
must be contrasted with the case of cuprates, where the density of
electrons $n$ is large (of order of $n\sim 1-x$, where $x$ is the number of
doped holes/electrons), but nonetheless the superfluid density $n_s$ is
small, and scales approximately with $x$, so that $J_s\sim xt$, where $t$
is a typical hopping energy scale.\cite{lee_review} For this reason, the
validity of the well-known Uemura plot,\cite{lee_review} i.e. the scaling
of $T_c$ with $J_s$ instead of the gap value $\D$ in underdoped cuprates,
does {\em not} signal any analogy between the two classes of
materials. Indeed, in pnictides $T_c\propto J_s\propto \D$, with a small
$J_s$ simply due to the fact that Fermi energy is small, in cuprates the
suppression of $J_s$ with respect to $\D$ is due to the proximity to the
Mott insulator, and the scaling of $T_c$ with $J_s$ can suggest a
predominant role of phase fluctuations.\cite{lee_review}

\section{Conclusions}
In the present work we propose an intermediate-coupling Eliashberg
multiband approach as an appropriate description of low-energy properties
of pnictides. As a starting point we use the bands measured by ARPES, where
an overall factor two of renormalization of the bands with respect to LDA
is observed,\cite{ding2,lu,yang,yi} which originate from correlations and
cannot be described by the coupling to a low-energy bosonic mode
($\o_0\simeq 20$ meV) discussed here within the Eliashberg theory. We focus
in particular on BKFA systems, where the multiband structure is accompanied
by a pronounced anisotropy of the Fermi-pocket sizes of the hole bands,
with an inner hole pocket almost nested to the electron one trough the
antiferromagnetic ${\bf Q}$ vector of spin fluctuations.  We have shown
that the simultaneous observation of two similar gap values in these bands
suggests that the predominant pairing is an interband one, as mediated by
the spin fluctuations between the set of hole pockets and the set of
electron pockets.  By comparing the calculations of the gaps with the
experimental data we have estimated the value of the interband coupling,
and we calculated the corresponding low-energy renormalization in several
spectral and thermodynamical properties. In particular, we showed that a
single set of parameter values can explain in a consistent way the data on
the specific heat\cite{mu} and on the superfluid density,
\cite{hiraishi,li} and we can predict the appearance of low-energy kinks in
the band dispersions, that are not always resolved in the
experiments. Consistently with the mean-field character of Eliashberg
theory we overestimate the critical temperature, which in real systems is
reduced by superconducting fluctuations,\cite{varlamov} whose relevance has
been recently pointed out in the context of paraconductivity
measurements.\cite{benfatto_fluct} Our analysis questions the possibility
of an extreme strong-coupling estimate as the one proposed recently in
Ref.\ [\onlinecite{umma}], because in this case one would find a huge mass
renormalization at low energy, that is in disagreement with the
experimental measurements of various thermodynamic quantities. Moreover, we
have clarified that one must resort to a full numerical Eliashberg
calculation, an issue that has been overlooked in previous studies of
multiband models with dominant interband interactions.\cite{dolgov} Indeed
we have explicitly shown that in this case the McMillan-Allen-Dynes-like
approximate expansion\cite{carbotte} fails already at coupling values
$\lambda\gtrsim 0.2 - 0.3$, well below the ones relevant for pnictides.

While the absolute values of the spectral and thermodynamic properties can
be captured only within a four-band Eliashberg theory, we have shown that
the temperature dependence of the same quantities do not show significant
deviations with respect to a two-bands BCS-like behavior, once the
renormalized parameters are used. This explains the success of many
two-band BCS fits proposed in the literature to reproduce the experimental
data. However, these fits must not be taken as indicative of the success of
a two-band BCS theory, that would instead completely fails both from the
qualitative and quantitative point of view in explaining the physics of
pnictides. 

While in the present work we focused on BKFA compounds, our results can be
extended to other families of pnictides, once that the proper modifications
due to the different nesting properties of the various Fermi pockets are
taken into account. An interesting example is provided by recent ARPES
reports in electron-doped BaFe$_{1.85}$Co$_{0.15}$As$_{2}$ ($T_c=25.5$ K),
where around the $\G$ point only the $\beta$ band crosses the Fermi level,
and $\D_\beta=6.6$ and $\D_\g=5$ meV.\cite{terashima_Co} In this case, the
almost nested bands are the $\beta$ pocket and the electron pockets
$\g_i$. However, if the bands have the same DOS measured in hole-doped
compounds and reported in Table I, the considerable DOS anisotropy between
these bands can explain why, even in the presence of the moderately large
interband pairing suggested by the $\D_i/T_c$ values, the gaps still differ
by 30$\%$. Our multiband Eliashberg scheme, with nesting-modulated pairing
strength, seems a suitable approach to be used to explain the material- and
doping-dependent hierarchy of the gaps in these pnictides as well. However,
more experimental information on the high-energy band renormalization would
be required to get more quantitative results. Indeed, also in
BaFe$_{1.85}$Co$_{0.15}$As$_{2}$ one observes the existence of a mass
renormalization beyond LDA at energy scales much higher that the one where
spin fluctuations are active.\cite{terashima_Co} A quantitative estimate of
these effects would help comparing the overall mass renormalization with
recent measurements of the specific-heat in Co-doped
BFA,\cite{meingast_cm09} where $\gamma$ seems to be smaller than what found
in K-doped crystals.  Indeed, as we suggest in the present work, such a
comparison is crucial to elucidate the dichotomy between high-energy and
low-energy mass renormalization effects. Thus, further theoretical and
experimental investigation in this direction can certainly help
understanding the physics of superconducting pnictides.

\section{Acknowledgements}
We thank G.A.  Ummarino for useful discussions. 
This work has been supported in part by the Italian
MIUR under the project PRIN 2007FW3MJX.

\end{document}